\documentclass[preprint, 12pt]{elsarticle}
\usepackage[colorlinks=true,linkcolor=blue,allcolors=blue]{hyperref} 
\usepackage{siunitx}
\usepackage{graphicx}  
\usepackage{subfigure}
\usepackage{multirow}
\usepackage{tabularx}
\usepackage{fancyhdr}
\usepackage{longtable}
\usepackage{parskip}
\usepackage[T1]{fontenc}
\usepackage{dcolumn}   
\usepackage{gensymb}
\usepackage{bm}        
\usepackage{amsfonts}  
\usepackage{amsmath}   
\usepackage{amssymb}   
\usepackage{hhline}
\usepackage{xcolor}
\usepackage{ulem}
\usepackage{changes}
\usepackage{gensymb}
\usepackage{caption} 
\journal{Journal of Magnetism and Magnetic Materials}

\begin{document}

\title{A double-spiral spin ordering in the helimagnet YBaCuFeO$_{5}$}

\author[1,2]{Yu-Hui Liang}
\author[1,2]{Chun-Hao Lai}
\author[3]{Chin-Wei Wang}
\author[3]{Shinichiro Yano}
\author[4]{Daisuke Okuyama}
\author[5]{Taku J. Sato}
\author[6,7,8]{Yusuke Nambu}
\author[3]{Shih-Chang Weng}
\author[3]{Yen-Chung Lai}
\author[9,10,11]{Wei-Tin Chen}
\author[12]{Kirrily C. Rule\corref{cor1}}
\author[1,2]{Chao-Hung Du\corref{cor2}}

\cortext[cor1]{Corresponding author: kirrilyr@gmail.com}
\cortext[cor2]{Corresponding author: chd@gms.tku.edu.tw}

\affiliation[1]{
            organization={Department of Physics, Tamkang University},
            city={New Taipei City},
            postcode={251301},
            country={Taiwan}}
\affiliation[2]{
            organization={Center of Advanced Spectroscopy and Smart Inspection for Material Research, Tamkang University}, 
            city={New Taipei City},
            postcode={251301},
            country={Taiwan}}
\affiliation[3]{
            organization={National Synchrotron Radiation Research Center},
            city={Hsinchu},
            postcode={300092}, 
            country={Taiwan}}
\affiliation[4]{
            organization={Institute of Materials Structure Science (IMSS), High Energy Accelerator Research Organization (KEK)},
            city={Ibaraki},
            postcode={305-0801}, 
            country={Japan}}
\affiliation[5]{
            organization={Institute of Multidisciplinary Research for Advanced Materials, Tohoku University},
            city={Sendai},
            postcode={980-8577}, 
            country={Japan}}
\affiliation[6]{
            organization={Institute for Materials Research, Tohoku University},
            city={Sendai},
            postcode={980-8577}, 
            country={Japan}}
\affiliation[7]{
            organization={Organization for Advanced Studies, Tohoku University},
            city={Sendai},
            postcode={980-8577}, 
            country={Japan}}
\affiliation[8]{
            organization={FOREST, Japan Science and Technology Agency},
            city={Saitama},
            postcode={332-0012}, 
            country={Japan}}
\affiliation[9]{
            organization={Center for Condensed Matter Sciences, National Taiwan University},
            city={Taipei},
            postcode={10617}, 
            country={Taiwan}}
\affiliation[10]{
            organization={Center of Atomic Initiative for New Materials, National Taiwan University},
            city={Taipei},
            postcode={10617}, 
            country={Taiwan}}
\affiliation[11]{
            organization={Taiwan Consortium of Emergent Crystalline Materials, National Science and Technology Council},
            city={Taipei},
            postcode={10622}, 
            country={Taiwan}}
\affiliation[l2]{
            organization={Australian Center for Neutron Scattering, Australian Nuclear Science and Technology Organization},
            state={NSW},
            postcode={2232}, 
            country={Australia}}

\begin{abstract}
Materials with a spiral spin ordering always show a rich phase diagram and can be a playground for studying the exotic physical properties associated with spiral magnetism. Using neutron elastic and resonant x-ray scattering on a high-quality single crystal YBaCuFeO$_{5}$, we demonstrate YBaCuFeO$_{5}$ to be a helimagnet consisting of a double-spiral spin ordering. YBaCuFeO$_{5}$ undergoes a commensurate to incommensurate magnetic phase transition at {\it T}$_\textnormal{N2}$$\sim$ 175 K, and the incommensurate phase consists of two spin-ordered components. Both components have different periodicities but with the same propagating direction along the {\it c}-axis below \textit{T}$_\textnormal{N2}$. Using resonant x-ray scattering at the Fe and Cu \textit{K}-edges, we further demonstrate that both spiral spin orderings result from the Fe$^{3+}$ and Cu$^{2+}$, respectively,  forming a double-spiral spin ordering structure. This can be understood to be caused by the coupling between both sublattices of Fe$^{3+}$ and Cu$^{2+}$ with the atomic lattice.
\end{abstract}



\maketitle

\section{INTRODUCTION}

Spiral magnetic structures may result from incommensurate spin coupling in which spins have a varying twist angle with respect to neighbors along the wavevector direction. Such a unique arrangement of the spin moments not only induces an antiferromagnetic component but could also a ferromagnetic component perpendicular to the wavevector direction. The coupling between both magnetic components and the crystal lattice results in a variety of exotic physical properties, such as the multiferroics driven by the spiral magnetism \cite{chdR2, chdR2A}, unconventional superconductivity observed in the spiral magnets CrAs or MnP \cite{chdR3, chdR4}, skyrmion spin textures in the chiral magnets \cite{chdR5, chdR6}, or frustrated magnetism in the Kagome material YMn$_{6}$Sn$_{6}$ \cite{chdR7, RMn6Sn62020}. Among those helimagnets, some of them, such as CrAs and YMn$_{6}$Sn$_{6}$, show the coexistence of two incommensurate structures in certain temperature regions, forming a double-spiral magnetic structure which is rare and also responsible for the observed unusual physical properties. Since the unique magnetic structure, spiral magnets are practically useful for the manufacture of spintronics, and the study of spiral magnets has also attracted lots of attention in condensed matter physics. In this article, we report the confirmation of a double-spiral magnetic structure in a double-layered perovskite oxide YBaCuFeO$_{5}$.

YBaCuFeO$_5$ (YBCFO) is a double-layered perovskite, and was derived from the investigation of the similar perovskite, high-Tc superconductor YBa$_{2}$Cu$_{3}$O$_{7-\delta}$ (YBCO) by substituting Fe for Cu \cite{ERRAKHO1988}. The crystal structure of YBCFO is constructed by removing the CuO chain layers in YBCO \cite{YBCFOandYBCFO1993}, so YBCFO and YBCO share a structural similarity.  Instead of the superconductivity as observed in YBCO, YBCFO undergoes an intriguing magnetic structural transition at approximately 200 K where the interplay between the structure and magnetism leads to an incommensurate antiferromagnetic ground state. This material has garnered significant scientific interest due to the observation of electric polarization and spiral magnetic structures in powdered samples below the phase transition temperature. Structurally, YBCFO is characterized by an ordered arrangement of \textit{A}-site elements Y and Ba, forming a tetrahedral framework. Internally, each \textit{B}-site elements Fe and Cu combine with oxygen to create pyramid structures, FeO$_{5}$ and CuO$_{5}$. The non-magnetic \textit{A}-site ions Ba, reside at the apex of these pyramids, while Y occupies the base. Contrary to YBCO, the addition of Fe in the lattice suppresses any superconducting properties in YBCFO \cite{YBCFOdopedFe1897, YBCOdopedFe1988}. However, YBCFO exhibits many magneto-structural phase transitions leading to novel magnetic behaviors \cite{NPDrefine4model2015, chemdisorder2018}. Understanding the origin of these phases and the delicate balance between the structure and magnetism motivates this study.

By means of the magnetization measurements, YBCFO was reported to have a magnetic phase transition from the paramagnetic to antiferromagnetic states at \textit {T}$_\textnormal{N1}$ $\sim$ 450 K \cite{ERRAKHO1988, MJR1998}. Subsequent reports identified a commensurate to incommensurate magnetic transition at \textit {T}$_\textnormal{N2}$ \cite{Mombru300K500K}, ${\it i. e.,}$ forming a spiral magnetic structure below \textit {T}$_\textnormal{N2}$, and the transition temperature ${T}_{N2}$ can be enhanced through various heat treatment processes \cite{YBCFOtuningTn2016}. However, for YBCFO, it is worth noting that the reported works have not been entirely consistent with each other because the magnetic properties are highly dependent on the distribution of the Cu/Fe which can be tuned by the synthesis route \cite{CAIGNAERT1995, MJR1998, Mombru1998, NPDrefine4model2015, YBCFOtuningTn2016, chemdisorder2022, PRM2024}. To explore the physical properties and potential applications of YBCFO, it is essential to understand its crystal and magnetic structures. Using the modified floating-zone method \cite{LAIYBCFOfz2015}, we are able to grow high-quality single crystals as shown in appendix information figure \ref{Fig1}\hyperref[Fig1]{(a)}. The diffraction pattern shows distinct and well-defined diffraction peaks, signifying a high-quality single crystal. The magnetic properties of the crystal were characterized by the use of magnetization measurements. As shown in Figure \ref{Fig1}\hyperref[Fig1]{(b)}, it demonstrates that the crystal of YBCFO shows a pronounced magnetic phase transition as the magnetic field applied perpendicular to the \textit{c}-axis at \textit {T}$_\textnormal{N2}$ $\sim$ 175 K. However, when the field is applied parallel to the \textit{c}-axis, no significant phase transition behavior is observed,
suggesting that the magnetic moment of YBCFO is oriented in the \textit{ab}-plane \cite{YBCFOSC2024}. These results are also consistent with the previously reported data \cite{CHDu2017}.

For this scattering study, we aim to provide a more comprehensive depiction of the low-temperature magnetic structure of YBCFO. In Section 3, we present evidence supporting the existence of the "double-spiral" magnetic structure in YBCFO. The manifestation of a double-spiral magnetic arrangement has been identified in various materials, including compounds of transition metal pnictides such as MnP \cite{MnPFePCrAs1974, MnP2014}, FeP  \cite{MnPFePCrAs1974, FeP1971, FeP2020, FeP2022}, CrAs \cite{MnPFePCrAs1974, CrAs1969, CrAs2016, CrAs2018, CrAs2020, CrAs2022, CrAs2023}, as well as rare-earth compounds like RERhSi (RE = Er, Tb) \cite{RERhSi1985, RERhSi2002} and REMn$_{2}$O$_{5}$ \cite{REMn2O51988, REMn2O52005} and Kagome lattice system, denoted as RMn$_{6}$Sn$_{6}$ (R = Sc, Y, Lu, Er) \cite{RMn6Sn61996, RMn6Sn62009, RMn6Sn62020, RMn6Sn62024}. By contrast to those reported materials showing a double-spiral magnetic structure, YBCFO is a perovskite oxide and with a higher magnetic transition temperature, combining both neutron scattering and resonant x-ray scattering, we demonstrate that YBCFO possesses a double-spiral magnetic structure resulting from the ordering of Fe$^{3+}$ and Cu$^{2+}$ respectively.

\section{EXPERIMENT}
For this study, a high-quality single crystal sample (\textit{a} = \textit{b} = 3.869 \si{\angstrom}, \textit{c} = 7.646 \si{\angstrom}, $\alpha$ = $\beta$ = $\gamma$ = 90$\degree$) with a size of about (7 $\times$ 5 $\times$ 3 mm$^{3}$) was used. The neutron elastic scattering experiment was carried out by using a thermal-neutron triple-axis spectrometer on the experimental station Taipan, at ANSTO. The collimation was set to open-40'-40'-open for high q-resolution and the analyzer was set to the same wavelength $\lambda$ = 2.345 \si{\angstrom} as the monochromator to measure purely elastic scattering. The sample was aligned in the (\textit{H} \textit{H} \textit{L}) scattering plane.

For resonant x-ray scattering experiments, the crystal was cut to an appropriate size of $\sim$ 6 $\times$ 3 $\times$ 1.5 mm$^{3}$ and was orientated to contain a scattering plane ($H H L$). The experiments were carried out on both experimental stations of BL 3 of Photon Factory and TPS 09A1 of NSRRC respectively.  Both experimental stations provide linearly polarised x-rays, and measurements were focused on both Fe and Cu {\it K}-edges and performed with a horizontal scattering geometry. Scan through the {\it L}-direction of Bragg reflection (1 1 1) gives an FWHM of about 0.0002 r.l.u., which is also taken as the experimental resolution. The experiment was not performed at the $L_3$-edge of  Fe and Cu, despite the more pronounced resonant effects at this edge for 3\textit{d} transition metals. This is due to the exceptionally long wavelengths of the \textit{L}-edge makes the magnetic reflection going beyond the diffraction Edward circle.

\section{RESULTS AND DISCUSSION}
\subsection*{\large Neutron Scattering}

Figure \ref{Fig2} shows the linear scans of the magnetic reflection \textbf{q$_\textnormal{CM}$} = (0.5 0.5 0.5) along the \textit{L}-direction as a function of temperature, obtained using elastic neutron scattering at TAIPAN. Detailed temperature-dependent results are shown in figure \ref{Fig2}\hyperref[Fig2]{(a)}. Figure \ref{Fig2}\hyperref[Fig2]{(b)} displays a sharp and intense singlet reflection at \textit{L} = 0.5, indicating that the magnetic order corresponds to the commensurate (CM) phase. The diffraction peak, with a width of 0.0076 r.l.u., defines the resolution for analyzing the neutron scattering data. Upon cooling to 200 K (figure \ref{Fig2}\hyperref[Fig2]{(c)}), two new broad satellite reflections next to the CM peak appear and are identified to be the incommensurate (ICM) component and marked as ICM$^{-}$ for \textit{L} < 0.5 and ICM$^{+}$ for \textit{L} > 0.5, respectively.  At 170 K (figure \ref{Fig2}\hyperref[Fig2]{(d)}), both incommensurate satellites (ICM$^{-}$ and ICM$^{+}$) develop into four diffraction peaks, which are marked as ICM1$^{-}$, ICM2$^{-}$, ICM1$^{+}$, and ICM2$^{+}$. It reveals that there are two distinct incommensurate magnetic structures with different periods at low temperatures. A similar behavior was also observed at (0 0 1.5). Therefore, the ICM component is defined as the satellite peak, which may encompass the two satellites in a disordered state when they cannot be resolved. The dashed lines in the figures represent the fitted result. The incommensurability of ICM1 and ICM2 noted by $\delta$ shows a temperature dependence, which is displayed in figure \ref{Fig2}. The q-vectors of the ICM1 and ICM2 are denoted as \textbf{q$_\textnormal{ICM1}$} = (0.5 0.5 0.5±$\delta$$_\textnormal{ICM1}$), $\delta$$_\textnormal{ICM1}$ $\sim$ 0.021 and \textbf{q$_\textnormal{ICM2}$} = (0.5 0.5 0.5±$\delta$$_\textnormal{ICM2}$), $\delta$$_\textnormal{ICM2}$ $\sim$ 0.049, respectively at 170 K. While the three phases of CM, ICM1 and ICM2 coexist around the transition region, the intensity of the CM peak remains unchanged. This infers that, at 170K, the predominant magnetic structure is still primarily the CM phase. However, as the temperature drops to 160 K (figure \ref{Fig2}\hyperref[Fig2]{(e)}), the intensity of the CM peak diminishes rapidly, while the intensity of the ICM1 increases markedly, and there is also a slight increase in the intensity of the ICM2. As the temperature decreases from \textit {T}$_\textnormal{N2}$, to 110K (figure \ref{Fig2}\hyperref[Fig2]{(f)}), the CM peak is no longer observed. Furthermore, the spacing between the diffraction peaks of ICM1$^{-}$ and ICM1$^{+}$ increases, and a similar phenomenon is observed for ICM2. As a consequence, the spacing $\Delta$ between the diffraction peaks of ICM1 and ICM2, defined as $\Delta$ = $\delta$$_\textnormal{ICM2}$ - $\delta$$_\textnormal{ICM1}$, decreases as temperature decreases. At 10 K (figure \ref{Fig2}\hyperref[Fig2]{(g)}), the lowest temperature of the experiment, both ICM1 and ICM2 are still distinguishable but with a tendency to merge at lower temperatures.  The incommensurability of ICM1 and ICM2 are $\delta$$_\textnormal{ICM1}$ $\sim$ 0.090 and $\delta$$_\textnormal{ICM2}$ $\sim$ 0.100 respectively at 10 K.

The fitting results, as shown in figure \ref{Fig3}, provide a clearer insight into the details of the phase transition process. The peak widths of the reflections are extracted from the deconvolution of the reflection profile with a resolution limit obtained from the magnetic reflection (0.5 0.5 0.5) at \textit{T} = 250 K. Figure \ref{Fig3}\hyperref[Fig3]{(a)} illustrates the peak positions of the diffraction peaks along the {\it L}-direction as a function of temperature. Between 200 K and 160 K, the commensurate and incommensurate magnetic phases coexist, while both the q-value of ICMI and ICM2 do not display a monotonic change around the transition boundary. This suggests a strong coupling between the magnetic ordering with the lattice in the transition boundary. A similar behavior was also reported in YMn$_2$O$_5$ \cite{YMO2006} and TbMn$_2$O$_5$ \cite{TbMO2004}\cite{Okamoto2007}.
Below 150 K, the CM phase vanishes, and the incommensurability $\delta$ of ICM1 and ICM2 increase, respectively. It is noteworthy that ICM1 and ICM2 are getting closer to each other at low temperatures. This implies that the $\delta$ values of the ICM1 and ICM2 components vary with temperature in different manners, suggesting that the origins of ICM1 and ICM2 are distinct distinct. Such a splitting behavior was not observed at Bragg reflections from the nuclear structure, such as (006), (116), and (2 2 6) \cite{CHDu2017}. This indicates that the atomic structure does not undergo any changes below \textit {T}$_\textnormal{N2}$, suggesting that the two distinct periodic magnetic structures observed do not result from inhomogeneous domains within the sample.

Figure \ref{Fig3}\hyperref[Fig3]{(b)} shows the evolution of the integrated intensity of each CM, ICM1 and ICM2 as a function of temperature. Near the phase transition temperature \textit{T}$_\textnormal{N2}$, the intensity of the CM phase rapidly decreases, the intensity of the ICM1 sharply increases, while ICM2 exhibits a minor increase. Since the integrated intensity of the incommensurate phase correlates with the rotation angle $\phi$ of magnetic moments within a magnetic unit cell \cite{CHDu2017}, comparing the integrated intensity depicted in figure \ref{Fig3}\hyperref[Fig3]{(b)} with the simulated results as reported by Lai {\it et. al.} \cite{CHDu2017},  the rotation angle $\phi$ of the ICM phase is inferred to be away from 180$\degree$ around \textit{T}$_\textnormal{N2}$, which could produce the ferromagnetic components in the {\it ab}-plane and be responsible for the unusual magnetic behavior as reported \cite{PRRYBCFOHfield2022}.  

The variation of the peak width (full-width-half-maximum, FWHM) for ICM1 and ICM2 with temperature is illustrated in figure \ref{Fig3}\hyperref[Fig3]{(c)}, showing the ordering evolution of the ICM1 and ICM2. Near \textit{T}$_\textnormal{N2}$, when ICM1 and ICM2 are just forming, the peaks exhibit broader widths, indicating a relatively shorter correlation length. However, around 170 K, the FWHM of ICM1 rapidly decreases, suggesting that ICM1 forms a highly ordered magnetic structure. On the other hand, ICM2 approaches the sharp line widths indicative of long-range order only at 10 K. This suggests that both ordering components originate from different sub-magnetic lattices and both magnetic lattices of ICM1 and ICM2 are in a competitive state below \textit{T}$_\textnormal{N2}$.  As a result, ICM1 dominates the magnetic structural behavior below \textit{T}$_\textnormal{N2}$, leading to the observed convergence of ICM1 and ICM2 at lower temperatures in figure \ref{Fig3}\hyperref[Fig3]{(a)}.
Distinct incommensurability $\delta$ implies different periods of the spiral spin ordering of ICM1 and ICM2.  The rotation angle $\Phi$ of the magnetic moment between the neighboring magnetic unit cells can be determined from the incommensurability $\delta$, as shown in figure \ref{Fig3}\hyperref[Fig3]{(d)}. The rotation angles $\Phi$ of both ICM1 and ICM2 decrease as the temperature decreases. Even at the lowest measured temperature of 10 K, there is still a disparity in the rotation angles between ICM1 and ICM2, {\it i.e.}$\Phi$$_\textnormal{ICM1}$ $\sim$ 147$\degree$ and $\Phi$$_\textnormal{ICM2}$ $\sim$ 144$\degree$. Therefore, below \textit {T}$_\textnormal{N2}$, there are two distinct spiral magnetic structures, forming a double-spiral spin ordering structure. 
It is noteworthy that at low temperatures, the intensity of ICM1 consistently surpasses that of ICM2 by several times as shown in figure \ref{Fig3}\hyperref[Fig3]{(b)}. According to Hund's rule, 
the Fe$^{3+}$ has a high-spin state with \textit{S} = 5/2, and Cu$^{2+}$ with \textit{S} = 1/2. Consequently, the magnetic moment of Fe$^{3+}$ is greater than that of Cu$^{2+}$. Since the intensity of the magnetic diffraction is proportional to the square of the magnetic moments, based on these experimental findings, we, therefore, infer that the magnetic structure of ICM1 primarily arises from the Fe$^{3+}$ and ICM2 from Cu$^{2+}$. In order to clarify the origins of both spin-ordering compounds, resonant elastic x-ray scattering was performed.

\subsection*{\large Resonant Elastic X-ray Scattering}

The low-temperature spiral magnetic structure of YBCFO is indeed quite intricate, but neutron scattering experiments do not clearly distinguish both sublattices formed by ICM1 and ICM2 because Fe and Cu have similar neutron scattering cross-sections. However, this issue can be resolved by using resonant x-ray scattering. Resonant elastic x-ray scattering (REXS) possesses element-selective capability due to the different absorption energy edges of Fe and Cu and can, therefore, effectively distinguish between the signals of these two elements. Furthermore, REXS also offers the advantage of enhancing the signal of magnetic diffraction peaks. Without the resonant effect, the signal from magnetic diffraction peaks would be exceedingly weak due to the small cross-section of the magnetic moment with x-rays \cite{NiO1972, Xrayweak1985}.
Experimentally, to minimize background and obtain a stronger signal for magnetic diffraction, the scattering geometry was set to be in the horizontal plane with incident $\pi$-polarized x-rays \cite{SigmaandPi1991}.
Using neutron scattering, we have demonstrated the existence of commensurate and incommensurate magnetic structures, as shown in figure \ref{Fig3},  so the x-ray scattering experiment was used to identify the origin of the magnetic reflections.

Figure \ref{Fig4}\hyperref[Fig4]{(a)} shows scattering geometry and the x-ray absorption spectrum crossing the Fe {\it K}-edge at 300 K. The spectrum exhibits an intense main absorption edge at $\sim$ 7.130 keV resulting from the Fe 1\textit{s} to 4\textit{p} dipole transition ($\Delta$\textit{l} = ±1). In addition, there is also a weak pre-edge feature at $\sim$ 7.1133 keV, which originates from the quadrupole transition ($\Delta$\textit{l} = ±2) from Fe 1\textit{s} to 3\textit{d} orbitals. This pre-edge feature is sensitive to the spin ordering of 3\textit{d} transition metals \cite{F2O31992, F3O4XRS2005}.
Using x-ray absorption spectroscopy, M. K. Srivastava1 {\it et. al.} observed a similar behavior and reported that the absorption edges of Fe and Cu do not change at low temperatures \cite{PongXAS2019}. This indicates that the valence states of Fe and Cu are still Fe$^{3+}$ and Cu$^{2+}$ at low temperatures. The origin of the magnetic ordering was confirmed by performing the scans through the magnetic reflections along the {\it L}-direction as a function of x-ray energy. Figure \ref{Fig4}\hyperref[Fig4]{(b)} displays the scans taken at {\it T} = 300 K for the magnetic reflection \textbf{q$_\textnormal{CM}$} = (1.5 1.5 0.5). It is clear to see the resonant effect at {\it E} = 7.113 keV. A similar resonant behavior, as shown in figure \ref{AppF1}\hyperref[APPF1]{(b)}, was also observed on the beamline BL3A at the Photon Factory of KEK. In figure \ref{AppF1}\hyperref[APPF1]{(a)}, the absorption spectrum shows a weak pre-edge feature at the Fe {\it K}-edge. Performing the slice scans ($\theta$-2$\theta$ scans of the magnetic reflection (1.5 1.5 0.5)) as a function of x-ray energy, the reflection shows a significant resonance at the pre-edge ({\it E} = 7.1155 keV) with a peak width of about 2 eV as shown in figure \ref{AppF1}\hyperref[AppF1]{(b)}. It indicates that the CM phase aligns as anticipated with the arrangement of the magnetic moments of Fe$^{3+}$. Concerning Cu$^{2+}$, we failed to observe the resonant effect at the Cu \textit{K}-edge for this commensurate magnetic structure. This is consistent with expectation, given that the magnetic moment of Cu is smaller than that of Fe, making it challenging to detect the resonant effect at the Cu ${\it K}$-edge.

For further measurements of the incommensurate phase of the ICM1 component, to enhance the reflection intensity, the incident x-ray energy was fixed at {\it E} = 7.113 keV. As displayed in figure \ref{Fig4}\hyperref[Fig4]{(d)} $\sim$ \hyperref[Fig4]{(g)},  for comparison, the blue lines show the magnetic reflection peaks obtained by neutron scattering, and the red lines represent the x-ray diffraction data. At 170 K, a sharp diffraction peak was observed in the REXS results, corresponding to the CM phase identified in neutron results (figure \ref{Fig4}\hyperref[Fig4]{(d)}). Although the neutron scattering results, display diffraction peaks for the ICM1 and ICM2 components, the correlation length of the ICM1 and ICM2 components near \textit{T}$_\textnormal{N2}$ are short, presenting challenges for observation in x-ray experiments. Upon cooling to 150 K, as shown in figure \ref{Fig4}\hyperref[Fig4]{(e)},  the CM phase was not detected by x-rays. Instead, two satellite diffraction peaks were detected in accord with the ICM1 component as observed by neutron scattering.  Resonance experiments were also taken at the Cu$^{2+}$ \textit{K}-edge, but the diffraction peak of ICM2 remained undetectable. This evidences that the ICM2 component was not detectable by resonant x-ray scattering at Fe or Cu {\it K}-edges. The REXS experimental results from BL3, Photon Factory, and TPS 09A1 are consistent. On the other hand, as particularly noteworthy in figure \ref{AppF1}\hyperref[AppF1]{(e)} and \ref{AppF1}\hyperref[AppF1]{(f)}, we performed the absorption spectrum at both Fe and Cu {\it K}-edge by fixing the scattering condition at the ICM1 position (1.5 1.5 0.55). The signal of the pre-edge of Fe is enhanced, but it does not show the enhancement at Cu {\it K}-edge. Based on these results, it therefore confirms that ICM1 is exclusively associated with  Fe$^{3+}$.

At 100 K (figure \ref{Fig4}\hyperref[Fig4]{(f)}) and 10 K (figure \ref{Fig4}\hyperref[Fig4]{(g)}), the x-ray diffraction results show the similar behavior as that observed at 150 K, and the q-values of the magnetic diffraction peaks still align with the  ICM1 diffraction peak. At 10 K, in the neutron scattering results, the ICM1 and ICM2 diffraction peaks are very close to each other, while the x-ray results show only the diffraction peaks coinciding with the ICM1 compound. To confirm the correlation between the incommensurate magnetic ICM1 component and Fe$^{3+}$, energy-dependent measurements, {\it i.e.} by a series of $\theta$-2$\theta$ scans through the diffraction peak as a function of x-ray energy, were conducted at 50 K at the diffraction peak \textbf{q$_\textnormal{ICM1}$} = (1.5 1.5 0.415). As shown in figure \ref{Fig4}\hyperref[Fig4]{(c)}, the integrated intensity shows an about sixfold enhancement at {\it E} = 7.113 keV. Given the higher spatial resolution, x-ray scattering is ideal for probing the long-range-ordered lattice with a length of up to a few thousand angstroms. As shown in figure \ref{Fig4}, at 170 K, the CM phase has an FWHM (the full width at half maximum) extracted from the neutron diffraction peak of about 0.0077 r.l.u., but it reaches up to about 0.0008 r.l.u. by x-ray scattering. This gives the CM phase a correlation length $\xi$ of about 3000 \si{\angstrom} {($\xi = \frac{1}{\frac{\pi}{c} \textnormal{FWHM(r.l.u.)}}$). At 10 K, the ICM1 component has a width of about 0.097 r.l.u. as extracted from neutron scattering, while a width of  0.0053 r.l.u. by x-ray scattering. Since we conducted x-ray measurements using different instruments, we are therefore confident that our measurements accurately represent the results from the entire sample.

Thus far, it has been confirmed that the spiral magnetic structure of the ICM1 component at low temperatures is exclusively composed of long-range ordered arrangements of Fe$^{3+}$ magnetic moments. Regarding the association between ICM2 and Cu$^{2+}$ magnetic moments, according to the absorption spectrum experiment \cite{PongXAS2019}, there is no prominent characteristic peak in the pre-edge region around the Cu \textit{K}-edge for YBCFO. This implies that it is difficult to probe the magnetic reflections caused by Cu using resonant x-ray scattering at the Cu {\it K}-edge. This leads to two outcomes. Firstly, lacking resonant effects, the signal from the magnetic diffraction peaks, which already exhibit much lower intensity for ICM2 than ICM1, would become exceptionally weak when viewed with REXS. The absence of resonance would make the observation of the spiral magnetic structure of ICM2 exceedingly challenging. Secondly, without resonance, it is impossible to establish a direct correlation between ICM2 and the long-range ordered Cu$^{2+}$ magnetic moments. However, in YBCFO, only two elements possess magnetic moments: the Fe$^{3+}$ and Cu$^{2+}$. It has been established that ICM1 is composed of Fe$^{3+}$ magnetic moments; it is, therefore, reasonable to infer that ICM2 is formed by the Cu$^{2+}$ magnetic moments.

The observation of two periodic magnetic structures from neutron scattering may arise from three possible scenarios: firstly, it could be questioned whether both incommensurate reflections result from two different domains in the crystal with different Fe/Cu ratios. If this is the case, since the transition temperature for the incommensurate phase is very sensitive to the Fe/Cu ratio \cite{YBCFOtuningTn2016,chemdisorder2018,chemdisorder2022,YBCFOLai2024}, both incommensurate magnetic structures of ICM1 and ICM2 should have different transition temperatures. However, according to the magnetization and neutron scattering data, as shown in figures \ref{Fig1}\hyperref[Fig1]{(b)} and \ref{Fig3}, the crystal used for this study shows only a sharp transition at \textit{T}$_\textnormal{N2}$ $\sim$ 175 K, which evidences a good homogeneity of the crystal. Additionally, given the small spot size of x-ray beams, resonant x-ray scattering could probe the different domains of the crystal, which could also give the different results. Nevertheless, our resonant experiments, as shown in figures \ref{Fig4} and \ref{AppF1}, show the same results. Again, this evidences a good homogeneity of the crystal. Secondly, both incommensurate magnetic structures could be generated from the same element such as the case in RMn$_6$Sn$_6$ (R = rare earths) \cite{chdR7, RMn6Sn61996, RMn6Sn62009, RMn6Sn62020, RMn6Sn62024}. Since the REXS detects only the resonant signal from Fe$^{3+}$ at the ICM1 component, which also evidences that both incommensurate magnetic structures of ICM1 and ICM2 have the different origins. The third possible scenario is whether Fe and Cu are ordered or disordered arrangement in the crystal space. A recent report by Gupta \textit{et al.} \cite{PRM2024}, using Mössbauer spectroscopy, demonstrates that the Fe ions occupy one of the sites in pyramid cells to form the ordered structure. On the other hand, a chemical-disordered model between Fe and Cu \cite{NPDrefine4model2015, chemdisorder2018, chemdisorder2022, YBCFOSC2024} was proposed to explain the formation of the incommensurate magnetic phase. Indeed, a fully disordered model can not form an ordered structure, so a chemical disorder is likely to be local, which is also in agreement with the theoretical model proposed by \cite{PRX2018}.

The schematic views of such a double-spiral spin ordering are displayed in figure 5. As shown in figure \ref{Fig5}\hyperref[Fig5]{(a)} and \hyperref[Fig5]{(b)}, the pitch length (=$\frac{1}{0.5-\delta}$) is extracted from the incommensurability as shown in figure \ref{Fig3}\hyperref[Fig3]{(a)} and describes the ordering length of the magnetic moments of spiral magnetic structure in the crystal space. Since the incommensurability depends on temperature, consequently, it gives the double-spiral magnetic structure to have a temperature-dependent periodicity as shown in figure \ref{Fig5}\hyperref[Fig5]{(c),(d),(e)} for the temperature at \textit{T} = 10, 100, and 150 K. 

\section{CONCLUSION}

In conclusion, utilizing the neutron elastic scattering on a high-quality single crystal of YBaCuFeO$_{5}$, we report that YBCFO possesses two spiral spin orderings with different ordering periods along the {\it c}-axis, forming a double-spiral magnetic structure below \textit {T}$_\textnormal{N}$$_\textnormal{2}$. Further, through resonant elastic x-ray scattering experiments, we identify the origin of the double-spiral magnetic structure in which the ICM1 component results from the ordering of Fe$^{3+}$ and ICM2 is inferred from Cu$^{2+}$. The double-spiral magnetic structure contains two spiral-spin ordering components, implying it should have different kinetic energies for the dynamic behavior. Further inelastic neutron scattering studies are planned to investigate this.

Higher-temperature double-spiral magnetic structure materials such as YMn$_{6}$Sn$_{6}$ and ErMn$_{6}$Sn$_{6}$ exhibit extremely complex phase diagrams under applied magnetic fields, showcasing rich and diverse behaviors \cite{Y166Hfield2021, RMn6Sn62024}. The study using powder of YBCFO, published in 2022 \cite{PRRYBCFOHfield2022}, also demonstrates intricate behavior under an applied magnetic field. Investigating the behavior of the double-spiral magnetic structure of YBCFO under a uniaxial magnetic field is also of interest. In addition, since YBCFO has a good lattice match with high Tc superconductor YBa$_{2}$Cu$_{3}$O$_{7}$, it could also be a good candidate substrate material for growing YBCO thin films. Given the unique characteristics of the double-spiral spin ordering, YBCFO may provide a rich playground for studying the field effects (electric and magnetic fields) of spiral magnetism or proximity effects between superconducting and spiral magnetic states.

\appendix
\section*{Appendix}

The REXS experiment was carried out on the experimental station BL3 of the Photon Factory at KEK. We observed the same behavior as that on the station TPS 09A. As shown in Fig. \ref{AppF1}\hyperref[AppF1]{(a)}, the absorption spectrum shows a weak pre-edge feature at the Fe {\it K}-edge. Performing the slice scans ($\theta$-2$\theta$ scans of the magnetic reflection (1.5 1.5 0.5)) as a function of x-ray energy, the reflection shows a significant resonance at the pre-edge ({\it E} = 7.1155 keV) with a peak width of about 2 eV as shown in Fig. \ref{AppF1}\hyperref[AppF1]{(b)}. Fig. \ref{AppF1}\hyperref[AppF1]{(c)} and \ref{AppF1}\hyperref[AppF1]{(d)} display the linear scans, as shown in red lines, through the {\it L}-direction of (1.5 1.5 0.5) with x-ray energy of 7.1155 keV at \textit{T} = 180 K and 130 K, respectively. The blue curves are the neutron data for comparison. It is clear to see that the resonant effect of both CM and ICM1 reflections mainly results from the Fe$^{3+}$. This is also evidenced by the fixed-q energy scans. We performed the absorption spectrum at both Fe and Cu {\it K}-edge by fixing the scattering condition at the ICM1 position (1.5 1.5 0.55). As shown in Fig. \ref{AppF1}\hyperref[AppF1]{(e) and (f)}, the signal of the pre-edge of Fe is enhanced, but it does not show the enhancement at Cu {\it K}-edge. 

\section*{Acknowledgements}

CHD is grateful to NSTC of Taiwan for funding Grant Nos. 112-2112-M-032-015 and 113-2112-M-032-002. The experiment done at KEK was supported by the GIMRT office of Tohoku University through proposal number 2021112-CNKXX-0502. WTC also acknowledges the NSTC, for fundings 111-2112-M-002-044-MY3 and 112-2124-M-002-012, and Academia Sinica project number AS-iMATE-111-12. The authors are grateful to NSRRC for providing the travel grant to carry out neutron experiments at ANSTO and also thank NSRRC and ANSTO for the arrangements of the experimental beamtimes. The authors would like to thank the ACNS staff for their assistance during beam time proposals P3441, P3407, P4800, and P5215.

\bibliographystyle{elsarticle-num} 
\bibliography{ref}

\renewcommand{\thefigure}{1}
\begin{figure}[h]
\centering
\includegraphics[width=\columnwidth]{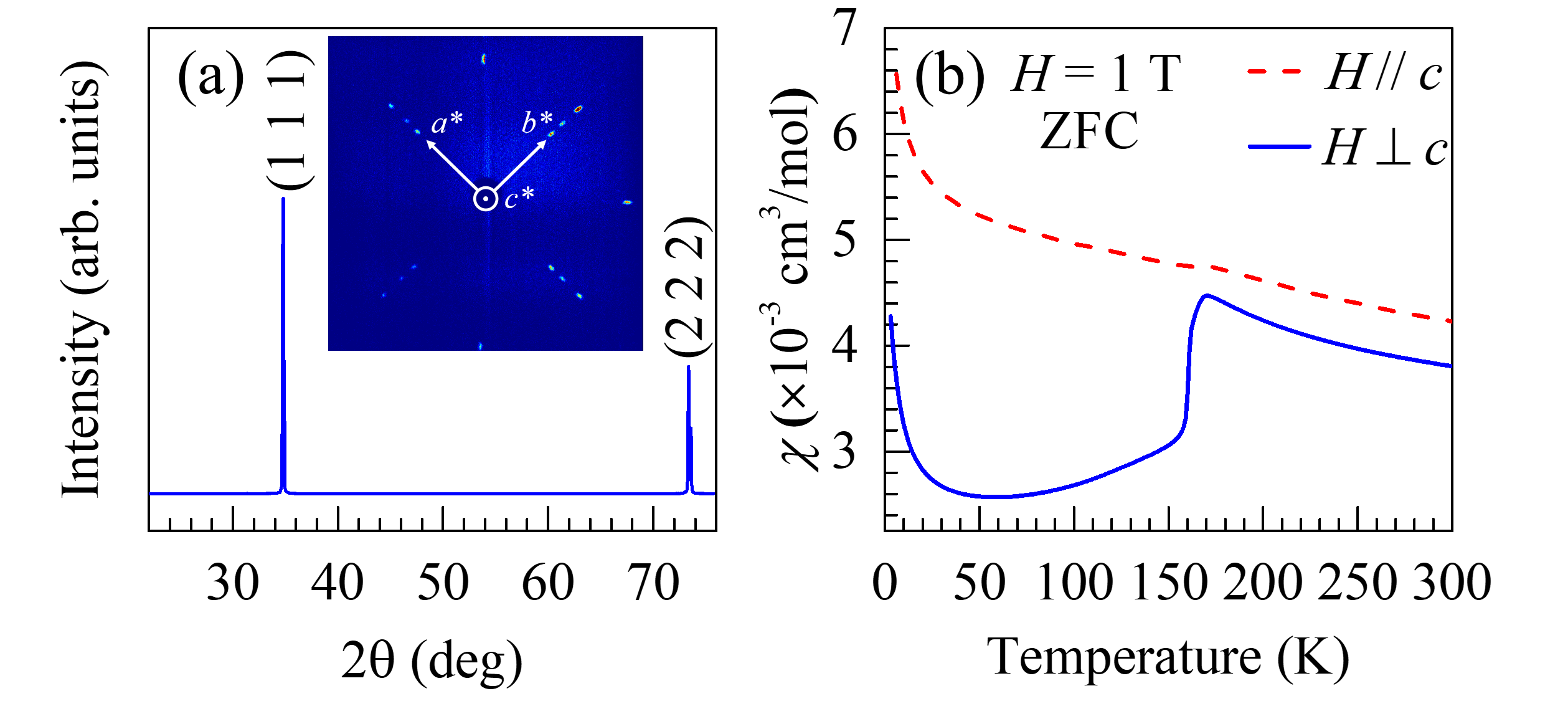}
\caption{(a) Diffraction pattern along the [1 1 1] direction of the single-crystal sample, obtained using an in-house four-circle diffractometer with Cu \textit{K}$_\alpha$ X-rays. The inset shows the Laue diffraction pattern taken with a Multiwire MWL-120, revealing distinct diffraction spots without powder rings. (b) The temperature-dependent magnetization of the single-crystal sample was measured using a Superconducting Quantum Interference Device (SQUID) magnetometer from Quantum Design. The blue solid line represents the magnetization perpendicular to the \textit{c}-axis, while the red dashed line corresponds to the magnetization parallel to the \textit{c}-axis.}
\label{Fig1}
\end{figure}

\renewcommand{\thefigure}{2}
\begin{figure}[h]
\centering
\includegraphics[width=\columnwidth]{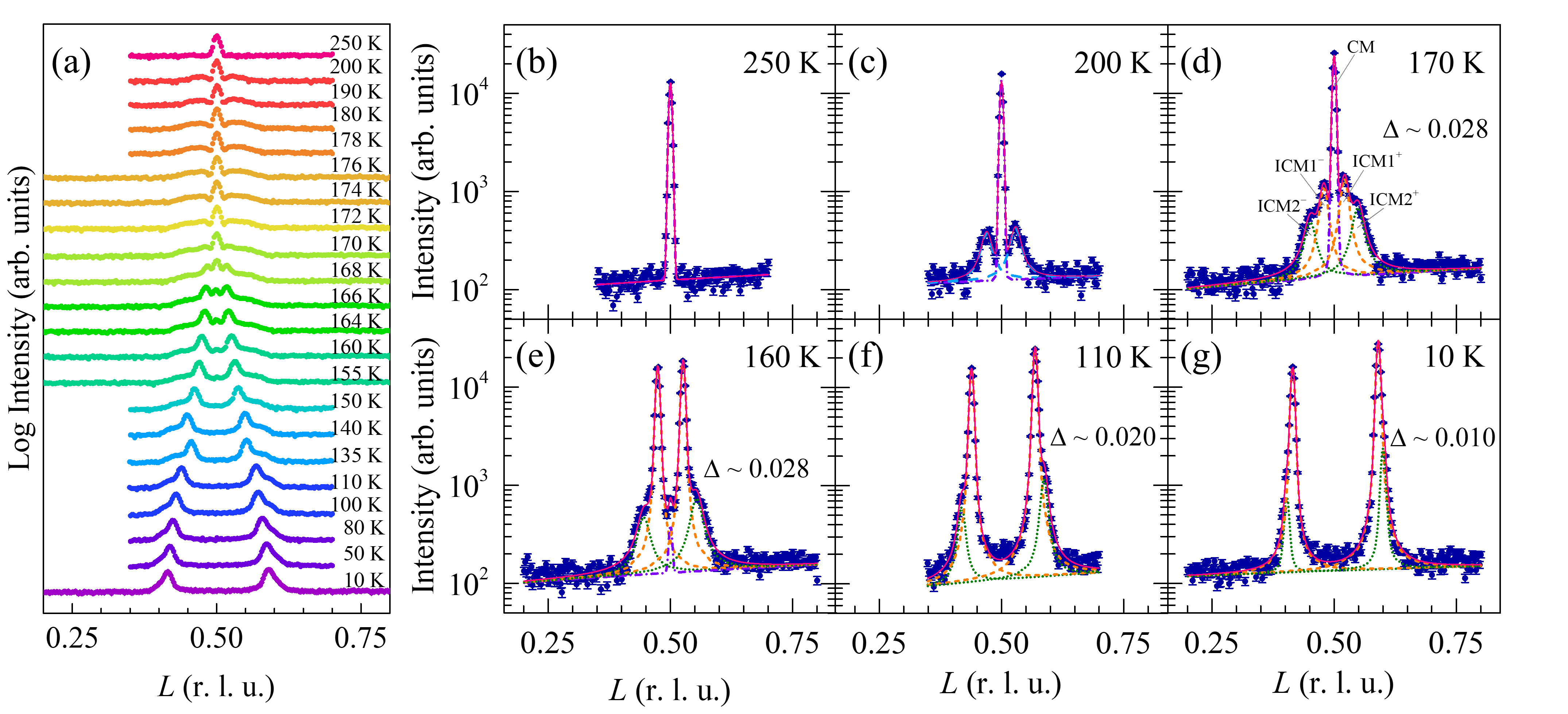} 
\caption{(a) The temperature-dependent neutron scattering data along the \textit{L}-direction of the magnetic reflection \textbf{q} = (0.5 0.5 \textit{L}), and at specific temperatures (b) 250 K, (c) 200 K, (d) 170 K, (e) 160 K, (f) 110 K, and (g) 10 K. The purple dashed dot lines, as shown in (b)$\sim$(e), represent the fitted peak for the CM phase, and the blue dashed lines shown in (c) correspond to the ICM component. The orange short dashed lines are for the ICM1, and the green short dot lines are for the ICM2 components, respectively, as shown in (d)$\sim$(g). And the $\Delta$ shown in (d)$\sim$(g) represents the difference in incommensurability between ICM1 and ICM2 ($\Delta = \delta_\textnormal{ICM2} - \delta_\textnormal{ICM1}$).}
\label{Fig2}
\end{figure}

\renewcommand{\thefigure}{3}
\begin{figure}[h]
\centering
\includegraphics[width=\columnwidth]{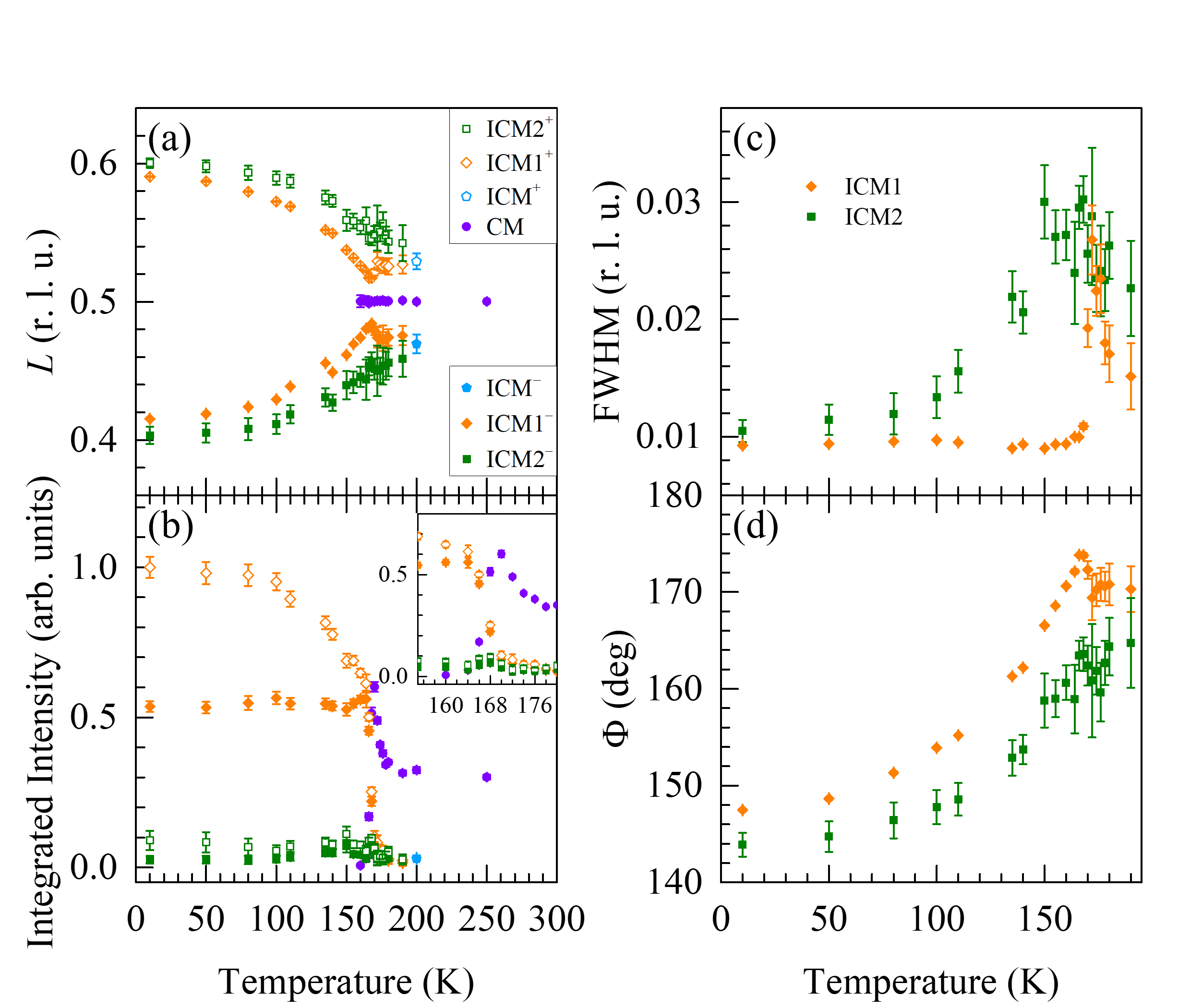} 
\caption{The results of fitted magnetic reflections (0.5 0.5 {\it L}) as a function of temperature.  (a) The temperature-dependent q-values along the \textit{L}-direction and (b) the integrated intensity of diffraction peaks. The purple points represent the CM phase, blue pentagonal points represent the ICM component, orange diamond points represent the ICM1 component, and green square points represent the ICM2 component. The superscript "-" denotes \textit{L} < 0.5, while "+" indicates \textit{L} > 0.5. (c) The FWHM of the diffraction peaks and (d) the variation of the rotation angle $\Phi$ between the neighboring magnetic unit cells for the spiral structure ICM1 plotted in orange circles, and ICM2 plotted in green circles, as a function of temperature.}
\label{Fig3}
\end{figure}

\renewcommand{\thefigure}{4}
\begin{figure}[h]
\centering
\includegraphics[width=0.7\columnwidth]{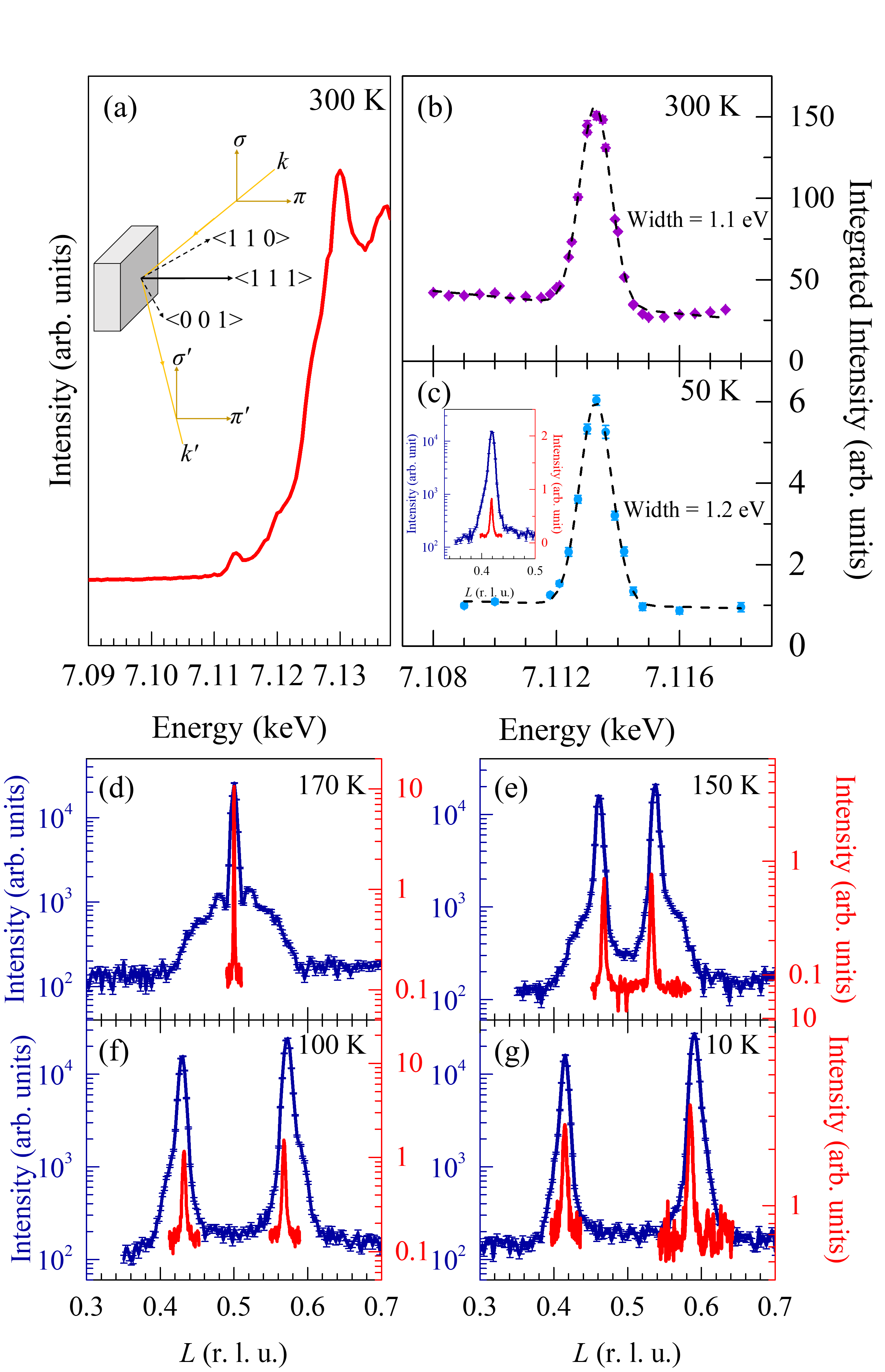}
\caption{Resonant elastic x-ray scattering (REXS) along the {\it L}-direction of the magnetic reflections (1.5 1.5 {\it L}) at different temperatures. (a) The absorption spectrum at 300 K for the Fe \textit{K}-edge, with a pre-edge excitation at 7.1133 keV, and the inset depicts a schematic representation of the scattering plane of the experiment. (b) and (c) The integrated intensity of the slice scans as a function of x-ray energy. It shows a resonant effect at the pre-edge \textit{E} = 7.113 keV at \textit{T} = 300 K and 50 K, respectively. The inset of (c) depicts magnetic reflection from neutron scattering (royal-blue) and resonant x-ray scattering (red) at a temperature of 50 K. (d)$\sim$(g) comparison of neutron and REXS data at 170 K, 150 K, 100 K, and 10 K for the magnetic reflection. The royal blue line represents neutron data, and REXS is represented by the red line. In (e), the orange dashed line at 150 K represents the diffraction pattern measured at \textit{E} = 8.990 keV.}
\label{Fig4}
\end{figure}

\renewcommand{\thefigure}{5}
\begin{figure}[h]
\centering
\includegraphics[width=\columnwidth]{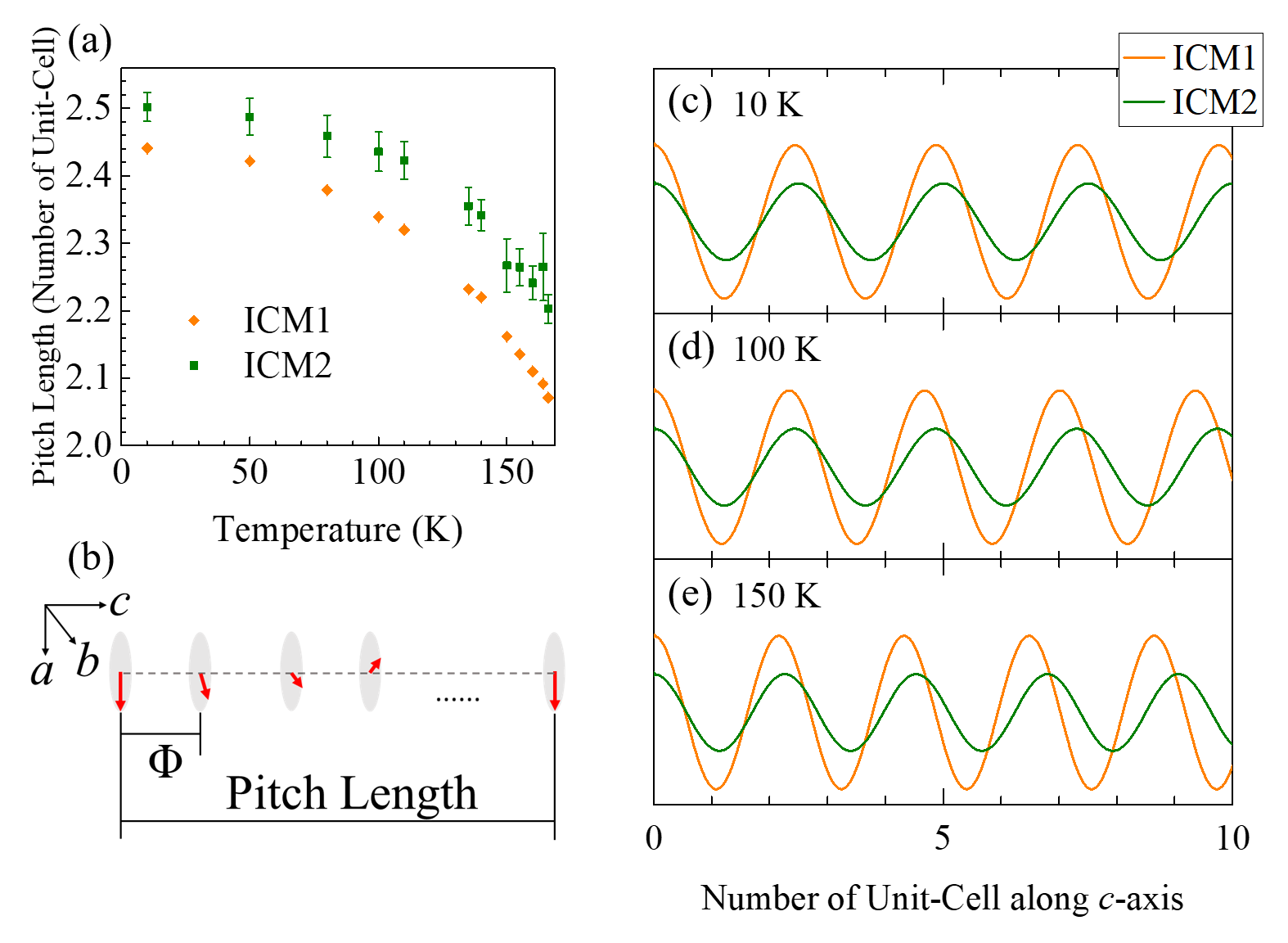}
\caption{(a) The pitch length of the ICM1 and ICM2 components as a function of temperature. (b) A schematic diagram of the rotation angles $\Phi$ and pitch length. Schematic diagram of the double-spiral magnetic structure at (c) 10 K, (d) 100 K, and (e) 150 K.}
\label{Fig5}
\end{figure}

\renewcommand{\thefigure}{A1}
\begin{figure}[h]
\centering
\includegraphics[width=\columnwidth]{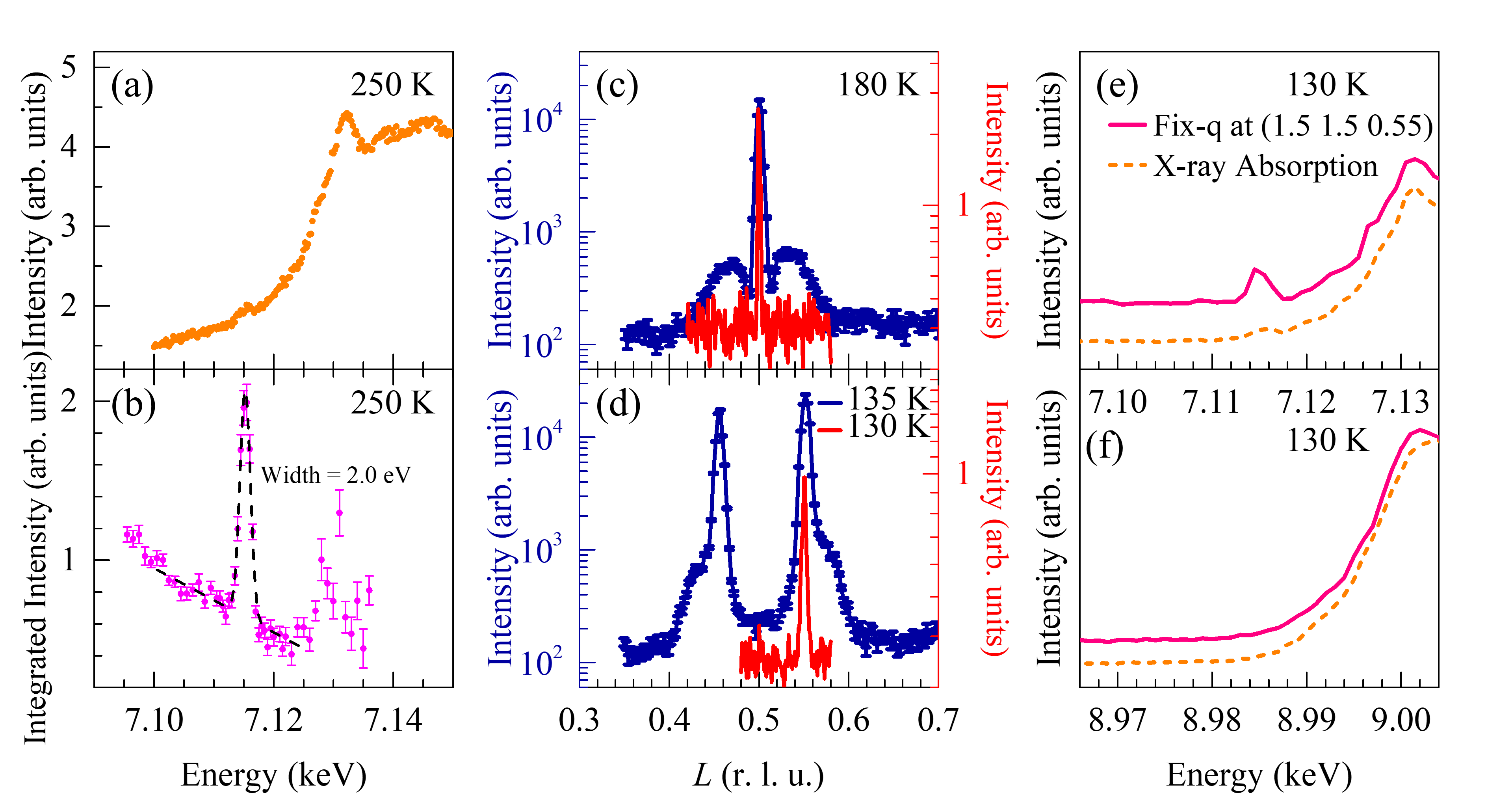}
\caption{Resonant x-ray scattering data obtained from the experimental station BL03 of Photon Factory. (a) Absorption spectrum at \textit{T} = 250 K for the Fe \textit{K}-edge, a weak pre-edge feature was observed at {\it E} = 7.1155 keV. (b) The integrated intensity of the magnetic reflection (1.5 1.5 0.5) versus the x-ray energy. The data were extracted from the $\theta$-2$\theta$ scans as a function of x-ray energy. (c) and (d) The linear scans through the {\it L}-direction of the CM reflection at \textit{T} = 180 K and ICM1 phase at \textit{T} = 130 K, respectively.
(e) and (f) Comparison of the absorption spectra at both Fe and Cu {\it K}-edge at \textit{T} = 130 K. The pink solid lines were obtained by fixing the scattering condition at the ICM1 position (1.5 1.5 0.55), and the orange dotted lines were obtained by off the detector away from the Bragg condition.}
\label{AppF1}
\end{figure}

\end{document}